# Analytical Study of a Monolayered Vibroacoustic Metamaterial

Majdi O. Gzal[a], Lawrence A. Bergman[b], Kathryn H. Matlack[a], Alexander F. Vakakis[a]

[a] Department of Mechanical Science and Engineering, University of Illinois, Urbana, USA

[b] Department of Aerospace Engineering, University of Illinois, Urbana, USA

**Abstract**

This study investigates a vibroacoustic phononic metamaterial system composed of repeated monolayered membrane-air cavity unit-cells to assess its efficacy in controlling sound waves. Assuming low-frequency axisymmetric modes, the coupled membrane-cavity vibroacoustic system for a representative unit-cell is solved entirely analytically. Unlike previous research that relied on an infinite series of eigenfunctions, our analysis offers a single-term exact solution for the membrane's displacement field, fully accounting for coupling with the acoustic cavities. Utilizing the transfer matrix method and the Bloch-Floquet theorem, we offer a comprehensive analytical characterization of the band structure, including closed-form analytical expressions for determining the bounding frequencies of the bandgaps and the dispersion branches. Interaction between Bragg and local resonance bandgaps is examined by adjusting Bragg bandgap positions, with detailed mathematical descriptions provided for their overlapping and transition. Additionally, a "plasma bandgap" analogous to metallic plasma oscillations is identified, with a derived analytical expression for its frequency. First two passbands remain robust against cavity depth variations, limiting wave manipulation capabilities. Analysis of the finite phononic system involves constructing the global transfer matrix to study natural frequencies and scattering coefficients. Interaction between Bragg and local resonance bandgaps in finite systems results in ultra-narrow passbands, creating transparency windows analogous to electromagnetically induced transparency by quantum interference. This theoretical framework enables precise characterization and engineering of bandgaps in the monolayered vibroacoustic phononic metamaterial, highlighting its potential for controlling low-frequency sound wave propagation across multiple frequencies.

**Keywords**

2D inhomogeneous wave equation; cylindrical acoustic cavity; circular elastic membrane; transfer matrix; vibroacoustic phononic metamaterial; plasma bandgap; Bragg bandgap; local resonance bandgap.

# 1. Introduction

Oscillating waves exhibit some similar characteristics irrespective of their type, i.e., longitudinal or transverse, elastic, acoustic, or electric. In classical mechanics, wave motion is classified into traveling waves, which can have variable frequency, and standing waves, which typically occur at permissible discrete frequencies. This classification extends to quantum mechanics as scattering and bound states. Moreover, periodically structured (phononic) media with an infinite number of unit-cells yield a third category of wave motion, characterized by continuous frequency bands were waves can propagate in space (pass bands), separated by impermissible gaps with exponentially decaying waves in space (stop bands). Within this framework, vibroacoustic phononic systems entail the interaction between fluid and elastic solid components, with the elastic solid components undergoing mechanical vibrations (dynamics), and the fluid components waves in enclosed domains or cavities (acoustics).

The study of wave propagation in periodic structured (phononic) media is a longstanding classical subject in physics, with a rich historical lineage dating back to Sir Isaac Newton [1]. Significant contributions include Lord Kelvin's description of dispersion phenomena [2] and Lord Rayleigh's demonstration of frequency bandgaps [3]. Subsequently, Brillouin's seminal book [4] on periodic structures, provided an in-depth exploration of the early advancements in this domain. The solution for wave propagation and dispersion in infinite ordered lattices is derived using Bloch theory [4-6], which implies translational symmetry of the wave functions. Consequently, the analysis of the infinite phononic system can effectively be reduced to treatment of a single unit-cell. The analysis of the resulting dispersion relation reveals the existence of impermissible frequency bands or bandgaps, where wave propagation in the periodic lattice is prohibited, i.e., Bloch waves become evanescent; and permissible frequency bands or passbands corresponding to propagating waves in the transmission spectrum of the phononic system. Although the band structure typifies solid-state physics [5,7-8], it is also observed in electromagnetic waves, known as photonic crystals [9-11], as well as in acoustic and elastic waves, termed sonic crystals [12-20]. The concept of a classical sonic crystal was first introduced to the public in Madrid [21], showcased as a generalization of the photonic crystal through a large periodic array of cylinders presented as a sound sculpture.

Due to the existence of bandgaps, the propagation of acoustic waves in a phononic medium can exhibit strong dispersion or complete reflection for frequencies within them. Consequently, these metamaterials provide unique capabilities for controlling and manipulating the wave-field, enabling the realization of novel wave phenomena and functionalities not found in natural materials. Within the realm of acoustics [22], notable achievements include the attainment of negative effective mass density [23,24], negative effective bulk modulus [25,26], simultaneously negative effective density and modulus (termed as double negativity (DNG)) [27,28], acoustic lenses for sub-diffraction imaging [29-32], and acoustic cloaking, rendering objects acoustically invisible [33,34].

A bandgap formation mechanism that naturally arises in phononic crystals is Bragg scattering, stemming from the periodicity of the structure. It arises from destructive interference between reflected and incident waves at successive unit-cell interfaces, necessitating wavelengths comparable to the unit-cell size [12,13]. This typically occurs at relatively higher frequencies, limiting low-frequency applications due to strict size requirements for engineering structures. As an attempt to avoid the limitations associated with Bragg bandgaps and achieve highly effective low-frequency attenuation with subwavelength unit-cell lengths, Liu et al. [35] introduced the concept of local resonance bandgap formation mechanism. These bandgaps stem from the interaction between the wave field and local resonances within the phononic crystal. Unlike Bragg bandgaps, the formation location of local resonance bandgaps is not dependent on structural periodicity and can occur at much lower frequencies. However, these resonance bandgaps tend to be narrow when the natural frequency of the unit-cell is significantly below the Bragg frequency.

As a promising solution to enhance the effective attenuation bandgap achieved by local resonance, several studies [36-47] have revealed that the interaction between Bragg scattering and local resonance bandgaps may yield enlargement of bandgap widths. Besides effectively widening the bandgaps, Yu and Wang [42] showed that acoustical transparency phenomena can be induced by the interplay of local resonance and Bragg bandgaps. However, in practical applications involving low-frequency wave manipulation (tailoring), the overlap between local resonance and Bragg frequencies, even the lowest one, is non-trivial and necessitates large periodicity for the structure. Additionally, suitable materials for achieving low-frequency manipulation, such as soft and heavy oscillators, may be impractical or challenging to implement. These factors collectively contribute to the complexity of tailoring acoustic waves at low frequencies, presenting significant challenges in engineering and scientific contexts. However, effective isolation of low-frequency sound is crucial due to the inherent difficulty in attenuating such sounds, as materials naturally possess weak dissipation properties within this frequency range. Moreover, low-frequency noise is widely recognized as a significant form of environmental pollution, primarily due to its high penetrating power, posing significant challenges to maintaining comfortable and healthy living and working environments [48,49].

There is considerable effort to develop diverse types of acoustic metamaterials to address the challenge of low-frequency sound attenuation, with a particular focus on thin, lightweight membrane-type acoustic metamaterials, which are capable of distinctive sound insulating capabilities. Indeed, due to the weak elastic moduli of the membranes, various low-frequency oscillation patterns can emerge, enhancing the potential for sound manipulation at lower frequencies. Membrane-type acoustic metamaterials have been applied as sound absorbers, leveraging their negative effective mass densities [23,24,50-56]. Yang et al. [23] have demonstrated experimentally that a circular elastic membrane with a small circular weight attached to its center can effectively attenuate sound transmission in the $100-1000\ Hz$ frequency range. Plasma-like behavior was experimentally observed in [24] where a finite one-dimensional (1D) phononic acoustic metamaterial with an array of elastic membranes was considered. In that study, the experimental findings showed that acoustic waves below the cut-off frequency $f_c = 735[Hz]$ were completely blocked; this experimental finding was explained using an approximate theory based on the effective density and bulk modulus of this phononic metamaterial. The performance and capabilities of membrane-type acoustic metamaterials in controlling and tailoring sound waves at low frequencies have been predominantly demonstrated through experiments and finite element analysis [50-56]. Analytical studies face unique challenges in analyzing such continuous vibroacoustic systems due to the high modal densities encountered, thus complicating the analysis and design. For instance, efforts have been made to calculate the sound transmission loss of membrane-type acoustic metamaterials analytically [55,56]. However, the analytical outcomes were obtained using the modal superposition method, yielding solutions in an infinite series form, thereby limiting the physical insight that can be gained. Additionally, analytical solutions for classical problems [57-60] on sound transmission through repetitive cells of coupled membranes and acoustic cavities, also led to solutions involving infinite series or integral forms which were of limited practical use.

In [61], a vibroacoustic phononic metamaterial was introduced to efficiently control longitudinal vibrations in ultra-low frequency ranges. The unit-cell of this metamaterial consisted of a single discrete mass-spring-damper system along with a cavity, offering only a single local resonance bandgap. This limitation restricted the practicality of that system for multi-frequency vibration absorption. In section 2 we introduce an infinite (i.e., composed of an infinite number of unit-cells) vibroacoustic metamaterial comprised of monolayered membrane-cavity unit-cells, leveraging the membrane's infinite local resonances and bandgaps to achieve *broadband performance*. Initially, assuming axisymmetry in the acoustics we employ a recently proposed approach [62] to develop exact single-term analytical solutions. This closed-form solution facilitates the construction of $(2 \times 2)$ local transfer and scattering matrices which enable exact Bloch – Floquet analysis of the system with infinite number of unit-cells. Then, in Section 3 the bounded system of N unit-cells is addressed by constructing a global transfer matrix employing the *Cayley–Hamilton* theorem [63-65]. The overall aim of this study is to develop a purely analytical methodology to comprehensively investigate and characterize the bandgap mechanisms within the proposed vibroacoustic metamaterial with a focus on understanding its capabilities and limitations for *low-frequency sound tailoring and control*. Finally, Section 4 presents concluding remarks and discusses limitations and potential applications of these findings.

## 2. Bloch analysis of the vibroacoustic metamaterial with infinite number of unit-cells

We consider a phononic vibroacoustic metamaterial in the form of a cylindrical duct containing an infinite set of unit-cells composed of monolayered membrane-cavity resonators (that is, there is a single membrane in each unit-cell). A schematic representation of a periodic assembly of three adjacent unit-cells is illustrated in Fig. 1a. This metamaterial consists of a succession of elastic circular membranes coupled by homogeneous acoustic air cavities of uniform depths $\Delta_1$. A typical unit-cell is shown in Fig. 1b. A local cylindrical coordinate system, represented by $(\hat{r}_{(n)}, \hat{\theta}_{(n)}, \hat{z}_{(n)})$, located at the center of the membrane, is considered for analyzing the $n^{th}$ unit-cell. The cylindrical duct of radius is assumed to be acoustically rigid in the radial direction, meaning that the radial component of the flow velocity at the duct's circumference is required to be zero. The duct of radius $a$ has the same radius as the (identical) membranes and cylindrical cavities. The density and sound speed of the air inside each cavity are denoted by $\rho_a$ and $c_a$, respectively. Each membrane is assumed to be clamped along its edge, with uniform thickness $h_m$, material density $\rho_m$, speed of elastic wave propagation $c_m$, and viscous damping $\lambda_m$. The thickness of the membrane is assumed to be much smaller in comparison to the cavity depth, i.e., $h_m \ll \Delta_1$; consequently, the characteristic length of the unit-cell, i.e., the spatial period of this phononic metamaterial, is simply equal to $\Delta_1$. In the following derivations, we will assume that the amplitudes of the membrane transverse vibrations are small enough for the linearized form of infinitesimal elasticity to apply. The interaction between adjacent unit-cells is described using the acoustic pressure and velocity, $p_{L_{(n)}}$ and $v_{L_{(n)}}$ at the left boundary ($L$), and by $p_{R_{(n)}}$ and $v_{R_{(n)}}$ at the right boundary ($R$), respectively. We assume that the acoustic pressure and velocity at the left boundary of the $n^{th}$ unit-cell are known, while the acoustic states (i.e., pressure and velocity) at the right boundary are unknown.

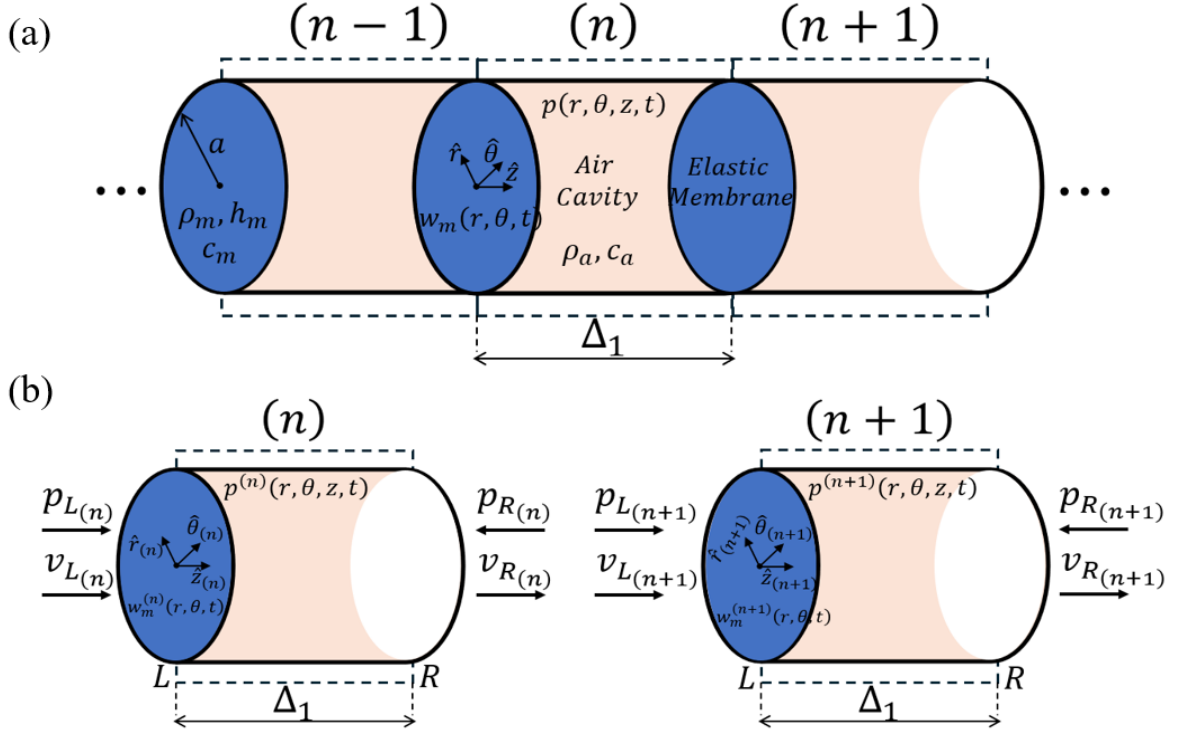

**Figure 1**: Schematic of the infinite vibroacoustic monolayered metamaterial: (a) Three adjacent unit-cells, and (b) the $n^{th}$ and $(n + 1)^{th}$ unit-cells showing membrane velocities and air pressures at their left ($L$) and right ($R$) boundaries; dashed frames mark the unit-cells, blue circle represents the membrane while light orange indicates the acoustic air cavity.

### 2.1. Local transfer and scattering matrices of the unit-cell

We start by deriving the transfer and scattering matrices for the single (say the $n^{th}$) unit-cell (see Fig. 1b), since, by imposing the Bloch-Floquet theorem, this will be the basis for studying the acoustics of the vibroacoustic phononic metamaterial of infinite axial extent. Let $p^{(n)}(r_{(n)}, \theta_{(n)}, z_{(n)}, t)$ denote the instantaneous acoustic pressure distribution within the cavity in that unit-cell. This pressure distribution is governed by the homogeneous 3D wave equation represented in cylindrical coordinates [57]:

$$c_a^2 \left( \frac{\partial^2 p^{(n)}}{\partial r_{(n)}^2} + \frac{1}{r_{(n)}} \frac{\partial p^{(n)}}{\partial r_{(n)}} + \frac{1}{r_{(n)}^2} \frac{\partial^2 p^{(n)}}{\partial \theta_{(n)}^2} + \frac{\partial^2 p^{(n)}}{\partial z_{(n)}^2} \right) = \frac{\partial^2 p^{(n)}}{\partial t^2} \qquad (1)$$

In addition, the acoustic pressure must satisfy the appropriate boundary conditions, which according to the linearized Euler transport equation are expressed as follows:

$$\left. \frac{\partial p^{(n)}}{\partial r_{(n)}} \right|_{r_{(n)}=a} = 0, \qquad \left. \frac{\partial p^{(n)}}{\partial z_{(n)}} \right|_{z_{1(n)}=0} = -\rho_a \frac{\partial^2 w_m^{(n)}}{\partial t^2}, \qquad (2)$$
$$\left. p^{(n)} \right|_{z_{(n)}=\Delta_1} = -p_{R_n}$$

Here, $w_m^{(n)}(r_{(n)}, \theta_{(n)}, t)$ represents the instantaneous transverse displacements of the membrane of the $n^{th}$ unit-cell. The forced damped transverse displacement distribution of the membrane is governed by the following inhomogeneous 2D wave equation represented in polar coordinates,

$$c_m^2 \left( \frac{\partial^2 w_m^{(n)}}{\partial r_{(n)}^2} + \frac{1}{r_{(n)}} \frac{\partial w_m^{(n)}}{\partial r_{(n)}} + \frac{1}{r_{(n)}^2} \frac{\partial^2 w_m^{(n)}}{\partial \theta_{(n)}^2} \right) + \frac{1}{\rho_m h_m} \left( p_{Ln} - p^{(n)}|_{z_{(n)}=0} \right)$$
$$= \frac{\partial^2 w_m^{(n)}}{\partial t^2} + \frac{\lambda_m}{\rho_m h_m} \frac{\partial w_m^{(n)}}{\partial t} \qquad (3)$$

with clamped boundary condition:
$$w_m^{(n)}\big|_{r_{(n)}=a} = 0 \qquad (4)$$

In this study, we consider relatively low frequency ranges and assume that the longitudinal sound modes of the air cavity, as well as the transverse modes of the flexible membrane are *axisymmetric*. Therefore, the acoustic pressure distribution inside the cavity and the transverse displacement of the membrane are independent of $\theta$, and we may set $\frac{\partial(\cdot)}{\partial \theta} = 0$.

For the air cavity, the steady-state axisymmetric acoustic pressure distribution with frequency $\Omega$ that satisfies the corresponding equation of motion (1), and the boundary conditions of a rigid duct circumference along the radial direction, is expressed as,

$$p^{(n)}(r_{(n)}, z_{(n)}, t) = \sum_{l=0}^{\infty} J_0(r_{(n)}\beta_l) \left( P_c^{(l)} \cos(z_{(n)}k_l) + P_s^{(l)} \sin(z_{(n)}k_l) \right) e^{i\Omega t} \qquad (5)$$

where $J_0$ represents the zeroth order Bessel function of the first kind, $P_c^{(l)}$ and $P_s^{(l)}$ denote the cosine and sine modal amplitudes, respectively, of the $l^{th}$ mode of $p^{(n)}$, $k_l$ is the $l^{th}$ mode wave number, and $\beta_l$ is the $l^{th}$ root of the first order Bessel function of the first kind divided by the membrane radius $a$, i.e., $J_1(a\beta_l) = 0$. The relationship between $k_l$ and $\beta_l$ is given by:

$$k_l = \sqrt{\frac{\Omega^2}{c_a^2} - \beta_l^2} \qquad (6)$$

The modal wave number $k_l$ can be real or imaginary. Real values correspond to propagating (volume) modes in the infinite phononic metamaterial, while imaginary values to localized (evanescent) modes which are spatially confined near each membrane. Therefore, the infinite acoustic pressure time series (5) can be truncated to a finite number of harmonics, denoted as $N_l$, based on the modal cut-on frequency $\Omega_l^{cut-on}$. This frequency determines the minimum threshold below which there is no propagation of the $l^{th}$ mode, and it is given by, $\Omega_l^{cut-on} = \beta_l c_a = \frac{\chi_l}{a} c_a$, where $\chi_l$ is the $l^{th}$ root of the first order Bessel function of the first kind. In this study, we will fix the air sound speed to $c_a = 343\ [m/s]$ and the membrane (or duct) radius to $a = 5\ [cm]$. Consequently, the cut-on frequency for the case of $l = 1$, indicating two modes, is equal to $\Omega_1^{cut-on} = 4{,}183\ [Hz]$. Therefore, in the frequency range of $0 < \Omega < \Omega_1^{cut-on} = 4{,}183\ [Hz]$, the infinite series of acoustic pressure distributions given in (5) can be truncated to a single harmonic with $l = 0$, representing *monoharmonic coupling*, as follows,

$$p^{(n)}(r_{(n)}, z_{(n)}, t) = \left( P_c^{(0)} \cos(z_{(n)}k_0) + P_s^{(0)} \sin(z_{(n)}k_0) \right) e^{i\Omega t} \qquad (7)$$

where $k_0 = \Omega/c_a$. This pressure distribution is in the form of uniform plane waves propagating in the axial z direction (see Fig. 1) and coupling the elastic membranes through a single harmonic. Accordingly, the acoustic pressure and velocity at the left and right boundaries of the $n^{th}$ unit-cell can be represented as plane waves (see Fig. 1b):

$$p_{L_{(n)}} = P_{Ln} e^{i\Omega t}, \quad v_{L_{(n)}} = V_{Ln} e^{i\Omega t} \qquad (8)$$

$$p_{R_{(n)}} = P_{R_n} e^{i\Omega t}, \quad v_{R_{(n)}} = V_{R_n} e^{i\Omega t}$$

where $i = (-1)^{1/2}$ is the imaginary constant. Note that in this transfer matrix formulation the quantities $P_{L_n}$ and $V_{L_n}$ are regarded as inputs, while $P_{R_n}$ and $V_{R_n}$ as outputs.

Introducing (7) and (8) into the equations of motion for the membrane (3) and applying asymmetry we obtain:

$$c_m^2 \left( \frac{\partial^2 w_m^{(n)}}{\partial r_{(n)}^2} + \frac{1}{r_{(n)}} \frac{\partial w_m^{(n)}}{\partial r_{(n)}} \right) + \frac{1}{\rho_m h_m} \left( P_{L_n} - P_c^{(0)} \right) e^{i\Omega t} = \frac{\partial^2 w_m^{(n)}}{\partial t^2} + \frac{\lambda_m}{\rho_m h_m} \frac{\partial w_m^{(n)}}{\partial t} \quad (9)$$

Rather than employing the traditional approach of representing the transverse displacement of the membrane through an infinite series of eigenfunctions that satisfy the clamped boundary conditions, this study adopts an alternative approach first introduced in a 1D system [62]. To this end, *the transverse displacements of the membrane are solved directly without employing an eigenfunction (modal) expansion*, considering that the forced steady-state transverse deformation of the membrane, being a linear time invariant continuum, supports a classical space-time separation. This procedure is followed by solving the nonhomogeneous ordinary differential equation as a superposition of a particular solution of the inhomogeneous equation and the homogeneous partial solution. Finally, the regularity conditions, i.e., $\left| w_m^{(n)} \right|_{r_{(n)}=0} < \infty$, and clamped boundary conditions, i.e., $w_m^{(n)} \big|_{r_{(n)}=a} = 0$, are applied. These steps result in the following exact, closed form (compact) solution for the transverse displacements of the forced damped membrane:

$$w_m^{(n)}(r_{(n)}, t) = \frac{\left( P_c^{(0)} - P_{L_n} \right)}{\eta_m^2 \rho_m h_m c_m^2} \left( 1 - \frac{J_0(r_{(n)} \eta_m)}{J_0(a \eta_m)} \right) e^{i\Omega t} \quad (10)$$

$$\eta_m = \sqrt{\frac{\Omega^2}{c_m^2} - \frac{i\Omega \lambda_m}{\rho_m h_m c_m^2}}$$

Hence, the transverse displacement distribution of the membrane is expressed in terms of the acoustic pressure inside the cavity, $P_c^{(0)}$, which is still unknown at this stage.

To fully characterize the vibroacoustic state of the $n^{th}$ unit-cell, one still needs to solve for the two pressure amplitudes $P_c^{(0)}$ and $P_s^{(0)}$, given in equation (7) and the acoustic pressure and velocity at the right boundary $P_{R_n}$ and $V_{R_n}$ in (8). To accomplish this, two equations are constructed by substituting (7), (8) and (10) into the two interface boundary conditions (2). Another two equations are obtained by utilizing the continuity of the normal acoustic velocity at the left and right edges of the unit-cell, which, based on the linearized Euler transport equation, can be expressed as:

$$-\frac{1}{\rho_a} \left[ \int \frac{\partial p^{(n)}}{\partial z_{(n)}} dt \right]_{z_{(n)}=0} = v_{L_n}, \quad -\frac{1}{\rho_a} \left[ \int \frac{\partial p^{(n)}}{\partial z_{(n)}} dt \right]_{z_{(n)}=\Delta_1} = v_{R_n} \quad (11)$$

After rather lengthy, but elementary, algebraic manipulations, the following expressions for the unknown pressure and velocity are obtained in terms of the acoustic pressure and velocity at the left boundary:

$$P_c^{(0)} = P_{Ln} + \frac{i}{\Omega}\sigma_m V_{Ln} \quad , \quad P_s^{(0)} = -\frac{i}{\Omega}\sigma_a V_{Ln}$$

$$P_{Rn} = -\cos(k_0\Delta_1)P_{Ln} + \frac{i}{\Omega}[\sin(k_0\Delta_1)\sigma_a - \cos(k_0\Delta_1)\sigma_m]V_{Ln}$$

$$V_{Rn} = \frac{\Omega}{i\sigma_a}\sin(k_0\Delta_1)P_{Ln} + \left[\cos(k_0\Delta_1) + \sin(k_0\Delta_1)\frac{\sigma_m}{\sigma_a}\right]V_{Ln} \tag{12}$$

where,

$$\sigma_m = \eta_m^2 \rho_m h_m c_m^2 \frac{J_0(a\eta_m)}{J_2(a\eta_m)}, \quad \sigma_a = \frac{\rho_a \Omega^2}{k_0} = \rho_a c_a \Omega$$

Finally, substituting expressions (12) into (7) and (10), the distributions of the pressure inside the cavity and the membrane transverse displacements become:

$$p^{(n)}(z_{(n)}, t) = \left(\left(P_{Ln} + \frac{i}{\Omega}\sigma_m V_{Ln}\right)\cos(z_{(n)}k_0) - \frac{i}{\Omega}\sigma_a V_{Ln}\sin(z_{(n)}k_0)\right)e^{i\Omega t} \tag{13}$$

$$w_m^{(n)}(r_{1(n)}, t) = \frac{i}{\Omega}V_{Ln}\left(\frac{J_0(a\eta_m) - J_0(r_{1(n)}\eta_m)}{J_2(a\eta_m)}\right)e^{i\Omega t}$$

As an interesting observation we note that when eliminating the effect of the cavity on the membrane, i.e., by substituting $P_{1c}^{(0)} = 0$ in equation (10), the *in-vacuo* membrane natural frequencies are obtained by solving the equation $J_0(a\eta_m) = 0$. However, as shown in (13), in the presence of the cavity, the corresponding natural frequencies of the membrane are the roots of $J_2(a\eta_m) = 0$.

The sound pressure at the left boundary of the membrane of the $n^{th}$ unit-cell, $p_{L_{(n)}}$ induces transverse vibrations of the membrane, and because of this interaction, part of the incident sound energy is reflected, part is dissipated by the membrane, and the remaining part is transmitted through the membrane creating sound waves in the air cavity. This vibroacoustic process can be characterized by the state vector $\mathbf{y}$ constructed from the acoustic pressure and the normal particle velocity distributions in the $z$ (axial) direction. Let $\mathbf{y}_{L_{(n)}}$ and $\mathbf{y}_{R_{(n)}}$ represent the acoustic state vectors at the left and right boundaries, respectively, of the $n^{th}$ unit-cell defined as,

$$\mathbf{y}_{L_{(n)}} = \begin{bmatrix} \frac{p_{L_{(n)}}}{\rho_a c_a} \\ v_{L_{(n)}} \end{bmatrix}, \quad \mathbf{y}_{R_{(n)}} = \begin{bmatrix} \frac{p_{R_{(n)}}}{\rho_a c_a} \\ v_{R_{(n)}} \end{bmatrix} \tag{14}$$

where normalized acoustic pressures are introduced. Velocity compatibility and the pressure equilibrium at the junction between the $n^{th}$ and the $(n+1)^{th}$ unit-cells (see Fig.1b), require that,

$$\mathbf{y}_{L_{(n+1)}} = \begin{bmatrix} \frac{p_{L_{(n+1)}}}{\rho_a c_a} \\ v_{L_{(n+1)}} \end{bmatrix} = \begin{bmatrix} -\frac{p_{R_{(n)}}}{\rho_a c_a} \\ v_{R_{(n)}} \end{bmatrix} \tag{15}$$

based on which we construct the *local transfer matrix* $\mathbf{T}$, which represents the recursion relation between the acoustic state vectors at the left boundaries of the $n^{th}$ and $(n+1)^{th}$ unit-cells as follows:

$$\mathbf{y}_{L_{(n+1)}} = \begin{bmatrix} \frac{p_{L_{(n+1)}}}{\rho_a c_a} \\ v_{L_{(n+1)}} \end{bmatrix} = \mathbf{T}\mathbf{y}_{L_{(n)}} = \mathbf{T}\begin{bmatrix} \frac{p_{L_{(n)}}}{\rho_a c_a} \\ v_{L_{(n)}} \end{bmatrix} \tag{16}$$

with

$$T = \begin{bmatrix} T_{11} & T_{12} \\ T_{21} & T_{22} \end{bmatrix} = \begin{bmatrix} \cos(k_0 \Delta_1) & -i\left(\sin(k_0 \Delta_1) - \cos(k_0 \Delta_1)\frac{\sigma_m}{\sigma_a}\right) \\ -i\sin(k_0 \Delta_1) & \cos(k_0 \Delta_1) + \sin(k_0 \Delta_1)\frac{\sigma_m}{\sigma_a} \end{bmatrix}$$

Note that the components of the local transfer matrix, $T$, are frequency dependent. Two successive unit-cells are coupled through the acoustic pressure and velocity inside the cavities, which are represented by a single harmonic. Therefore, the proposed lattice is created from *mono-coupled unit-cells*, the local transfer matrix $T$ is of dimensions (2 × 2).

As the considered system is linear and time-invariant the *reciprocity principle* holds [66], necessitating that $det(T) = T_{11}T_{22} - T_{12}T_{21} = 1$, i.e., $T$ is a unimodular matrix and invertible (non-singular) regardless of frequency $\Omega$. By construction, the matrix $T$ describes how the acoustic state evolves as a function of frequency from the left boundary of the $n^{th}$ unit-cell to the left boundary of the $(n+1)^{th}$ one. Moreover, the inverse local transfer matrix $T^{-1} = adj(T)$ can be obtained by interchanging the state vectors in equation (14) and corresponds to vibroacoustic transmission in the opposite direction (i.e., right to left) sharing the same eigenvalues and eigenvectors as $T$. Physically, the unimodularity of the transfer matrix in this reciprocal system implies identical vibroacoustic spectrum in both axial directions (i.e., left to right and right to left).

The acoustic pressure at the left and right sides of the $n^{th}$ unit-cell can be viewed as a superposition of two waves propagating in opposite directions as,

$$p_{L_{(n)}} = p^-_{L_{(n)}} + p^+_{L_{(n)}}$$
$$p_{R_{(n)}} = p^-_{R_{(n)}} + p^+_{R_{(n)}} \tag{17}$$

where $p^+_{L_{(n)}}$ and $p^+_{R_{(n)}}$ represent the amplitudes of plane waves propagating in the positive $z$ direction at the left and right sides, respectively, of the $n^{th}$ unit-cell. Similarly, $p^-_{L_{(n)}}$ and $p^-_{R_{(n)}}$ denote the amplitudes of plane waves propagating in the negative $z$ direction at the left and right sides, respectively, of the $n^{th}$ unit-cell. Therefore, the acoustic velocities at the boundaries of the $n^{th}$ unit-cell can be expressed in terms of the pressure drop and air impedance, as follows:

$$v_{L_{(n)}} = \frac{1}{\rho_a c_a}\left(p^+_{L_{(n)}} - p^-_{L_{(n)}}\right)$$
$$v_{R_{(n)}} = \frac{1}{\rho_a c_a}\left(p^+_{R_{(n)}} - p^-_{R_{(n)}}\right) \tag{18}$$

Lastly, the acoustical behavior of the infinite periodic lattice can be described completely by the *local scattering matrix* $S$, which relates the incoming pressure wave vector $\left[p^+_{L_{(n)}}, p^-_{R_{(n)}}\right]^T$ and outgoing pressure wave vector $\left[p^-_{L_{(n)}}, p^+_{R_{(n)}}\right]^T$ in terms of the components of the local transfer matrix, $T$,

$$\begin{bmatrix} p^-_{L_{(n)}} \\ p^+_{R_{(n)}} \end{bmatrix} = S \begin{bmatrix} p^+_{L_{(n)}} \\ p^-_{R_{(n)}} \end{bmatrix}$$

where,

$$S = \begin{bmatrix} S_{11} & S_{12} \\ S_{21} & S_{22} \end{bmatrix} = \begin{bmatrix} \dfrac{T_{11} + T_{12} - T_{21} - T_{22}}{T_{11} + T_{12} + T_{21} + T_{22}} & \dfrac{-2}{T_{11} + T_{12} + T_{21} + T_{22}} \\ \dfrac{2}{T_{11} + T_{12} + T_{21} + T_{22}} & -\dfrac{T_{11} - T_{12} + T_{21} - T_{22}}{T_{11} + T_{12} + T_{21} + T_{22}} \end{bmatrix} \tag{19}$$

The local scattering matrix, $\mathbf{S}$, completely describes the transmitted and reflected waves, and when losses are included, the energy dissipation in each unit-cell. Specifically, the diagonal elements $S_{11}$ and $S_{22}$ determine the reflection coefficients ($R^+$ and $R^-$) of the unit-cell excited by positive- and negative-going plane waves, respectively. The off-diagonal elements $S_{21}$ and $S_{12}$, on the other hand, denote the transmission coefficients ($T^+$ and $T^-$) of the unit-cell excited by positive- and negative-going plane waves, respectively. The superscripts $(+,-)$ indicate the direction of the incident wave. The exact relations are expressed as follows:

$$\begin{aligned} R^+(\Omega) &= |S_{11}|^2, \quad T^-(\Omega) = |S_{12}|^2 \\ T^+(\Omega) &= |S_{21}|^2, \quad R^-(\Omega) = |S_{22}|^2 \\ \alpha^+(\Omega) &= 1 - R^+ - T^+, \\ \alpha^-(\Omega) &= 1 - R^- - T^- \end{aligned} \qquad (20)$$

Here $\alpha^+$ and $\alpha^-$ denote the corresponding absorption coefficients indicating the extent of energy dissipation due to intrinsic losses in the unit-cell. In terms of the scattering coefficients, the reciprocal behavior of the system implies that $T^- = T^+$, i.e., the transmission does not depend on the direction of the incident wave. However, since the unit-cell is asymmetric, the reflection coefficients are different ($R^+ \neq R^-$), and therefore the absorption does depend on the direction of propagation.

## 2.2. Passbands and stopbands (bandgaps)

Studying wave propagation within the vibroacoustic phononic metametarial with infinite number of unit-cells is accomplished by the application of Bloch-Floquet theorem, seeking solutions where the state vectors of two adjacent unit-cells are related by the expression:

$$\mathbf{y}_{L(n+1)} = \Lambda \mathbf{y}_{L(n)} \qquad (21)$$

where $\Lambda$ is a matrix of Floquet multipliers. Incorporating the transfer matrix relation from (16) into (21) yields the following linear eigenvalue problem:

$$\mathbf{T}\mathbf{y}_{L(n)} = \Lambda \mathbf{y}_{L(n)} \qquad (22)$$

This relation shows that the eigenvalues of the $\mathbf{T}$ matrix, denoted as $\Lambda_1$ and $\Lambda_2$, represent the Floquet multipliers at a given frequency. The evolution of state vector amplitudes between two consecutive unit-cells is encoded in the eigenvalues $\Lambda_{1,2}$ of the transfer matrix $\mathbf{T}$, which are solutions of the characteristic polynomial $\Lambda^2 - tr(\mathbf{T})\Lambda + 1 = 0$. It follows that, due to the reciprocal nature of the considered vibroacoustic system, the eigenvalues of the transfer matrix are only determined by its trace:

$$tr(\mathbf{T}) = T_{11} + T_{22} = 2\cos(k_0\Delta_1) + \sin(k_0\Delta_1)\frac{\sigma_m}{\sigma_a} \qquad (23)$$

Let the $j^{th}$ eigenvalue of the $\mathbf{T}$ matrix be related to a corresponding *propagation constant* $\mu_j$, defined by $\Lambda_j = e^{i\mu_j}, j = 1,2$. In general, the propagation constants are complex numbers, with their real and imaginary parts being functions of frequency. Specifically, the real part denotes phase lag, while the imaginary part logarithmic decay of the state between adjacent unit-cells. It follows that real propagation constants correspond to waves that propagate unattenuated through the (infinite, boundless) metamaterial, whereas purely imaginary propagation constant to evanescent (near field) waves with exponentially decaying envelopes. Complex propagation constants signify propagating waves (or complex modes) but with exponentially decaying envelopes, and appear in systems with internal damping or in some phononic systems with multiple coupling coordinates between unit-cells.

As the local transfer matrix $\mathbf{T}$ is unimodular its eigenvalues form reciprocal pairs,

$$\Lambda_1 = e^{i\mu}, \; \Lambda_2 = e^{-i\mu} \tag{24}$$

and the corresponding eigenvectors $X_1$ and $X_2$, are expressed as:

$$X_1 = \begin{bmatrix} 1 \\ \dfrac{e^{i\mu} - T_{11}}{T_{12}} \end{bmatrix}, \quad X_2 = \begin{bmatrix} 1 \\ \dfrac{e^{-i\mu} - T_{11}}{T_{12}} \end{bmatrix} \tag{25}$$

Lastly, the propagation constant $\mu$ is directly obtained by using the first principal invariant, specifically, the trace of the matrix $\boldsymbol{T}$, as follows:

$$\mu(\Omega) = arccos\left(\frac{1}{2} tr(\boldsymbol{T})\right) \tag{26}$$

In the non-dissipative system, i.e., $\lambda_A = \lambda_B = 0$, wave propagation occurs when $\mu$ is real, and this holds when $-2 \leq tr(\boldsymbol{T}) \leq 2$. This defines the *passbands* of the infinite system, the corresponding eigenvalues of the transfer matrix satisfy $|\Lambda_{1,2}(\Omega)| = 1$, and the bounding frequencies $\Omega_b$ of the passbands are computed by the relation $tr(\boldsymbol{T})|_{\Omega=\Omega_b} = \pm 2$. Alternatively, wave attenuation occurs when $\mu = i\theta$ or $\mu = \pi + i\theta$, where $\theta$ is a real phase. The resulting *stopbands* or *bandgaps* are computed by requiring that $tr(\boldsymbol{T}) < -2$ or $tr(\boldsymbol{T}) > 2$, and then one of the eigenvalues satisfies $|\Lambda(\Omega)| > 1$ and the second $|\Lambda(\Omega)| < 1$. Finally, in the case of a dissipative system, both the real and imaginary parts of $\mu$ are non-zero, indicating the presence of complex modes. Given the unimodularity of the local transfer matrix $\boldsymbol{T}$, without loss of generality, the band structure of the infinite vibroacoustic metamaterial will be analyzed only for the case of negative-going waves, i.e., for positive $\mu$.

We denote by $\Omega_{b(n)}^{cut-on}$ and $\Omega_{b(n)}^{cut-off}$ the cut-on and cut-off frequencies of the $n^{th}$ propagation band, respectively, such that $\ldots < \Omega_{b(n)}^{cut-on} < \Omega_{b(n)}^{cut-off} < \Omega_{b(n+1)}^{cut-on} < \Omega_{b(n+1)}^{cut-on} < \cdots$. These are computed by solving the transcendental characteristic equation,

$$2cos(k_0 \Delta_1) + sin(k_0 \Delta_1) \frac{\sigma_m}{\sigma_a} = \pm 2 \tag{27}$$

where $k_0, \sigma_a$, and $\sigma_m$ are dependent on the frequency $\Omega$. It follows that the corresponding width of the $n^{th}$ bandgap, denoted by $W_{(n)}^{gap}$, is simply:

$$W_{(n)}^{gap} = \Omega_{b(n+1)}^{cut-on} - \Omega_{b(n)}^{cut-off} \tag{28}$$

Consequently, the analysis of the band structure of the infinite vibroacoustic phononic metamaterial is reduced to computing the roots of (27), which, in turn, enables the comprehensive understanding of the governing mechanisms of the vibroacoustic coupling interactions within the band gaps.

Referring to the characteristic equation (27), both cut-on and cut-off bounding frequencies for the $n^{th}$ bandgap generally demonstrate complex nonlinear dependencies on the cavity depths ($\Delta_1$), i.e., the periodicity of the system, and the air ($\sigma_a$), and membrane properties ($\sigma_m$). This computation can be simplified and split into the following canonical categories which define two different types of bandgaps,

$$\begin{aligned}
&\textit{Type I}: \quad sin\left(\frac{\Omega_b \Delta_1}{c_a}\right) = 0 \;\rightarrow\; \Omega_b = \frac{l\pi c_a}{\Delta_1}, l \in \mathbb{N} \\
&\textit{Type II}: \quad \begin{cases} cot\left(\dfrac{\Omega_b \Delta_1}{2c_a}\right) = -\dfrac{\sigma_m}{2\sigma_a} \\ tan\left(\dfrac{\Omega_b \Delta_1}{2c_a}\right) = \dfrac{\sigma_m}{2\sigma_a} \end{cases}
\end{aligned} \tag{29}$$

where (29) describe the cut-on and cut-off frequencies of the bandgaps.

*Type I bandgaps are directly associated with the standard Bragg scattering mechanism* due to the phononic (periodic) configuration of the metamaterial. This mechanism involves the destructive interference of waves scattered from periodic inhomogeneities with wavelengths comparable to the spatial periodicity of the phononic metamaterial. Denoting the wavelength by $\tilde{\lambda}$, the $l^{th}$ order simple Bragg-diffraction condition is given by $\tilde{\lambda} = \frac{2\Delta_1}{l}$, where $\tilde{\lambda} = \frac{2\pi}{k_0} = \frac{2\pi c_a}{\Omega}$; alternatively, Bragg diffraction occurs when an integer number of half-wavelengths fits into one unit-cell. Regardless the local resonator in the unit-cell, i.e., the membrane, the cut-off frequency of the Bragg bandgap is dependent on the periodicity of the system (in the monolayer case $\Delta_1$). In contrast, *Type II bandgaps are generated exclusively due to the fluid-structure interaction within the unit-cells*, and are exclusively due to the membranes, which can be regarded as internal resonators within each unit-cell. From a mechanics perspective, achieving these two types of passbands enables the design of vibroacoustic metamaterials for wave tailoring, even at low (and ultra-low) frequencies.

Throughout this study (unless stated otherwise), the following system parameters are selected,

$$c_a = 343 \ [m/s], \quad \rho_a = 1.255 \ [kg/m^3]$$
$$h_m = 0.0011 \ [m], \ \rho_m = 1050 \ [kg/m^3], \ a = 0.05 \ [m] \quad (30)$$
$$\lambda_m = \begin{cases} 0 & lossless \\ 30 & lossy \end{cases} [sN/m^3]$$

where we utilized the actual values for the air speed ($c_a$) and density ($\rho_a$), and selected Polydimethylsiloxane (PDMS) as the material for the membrane. The speed of elastic wave propagation for the membrane ($c_m$) and the depth of the cavity ($\Delta_1$) are used as design parameters to investigate the behavior of the band structure. In practice, controlling the membrane wave speed ($c_m$) with a given material and geometry is simply achieved by adjusting its tension.

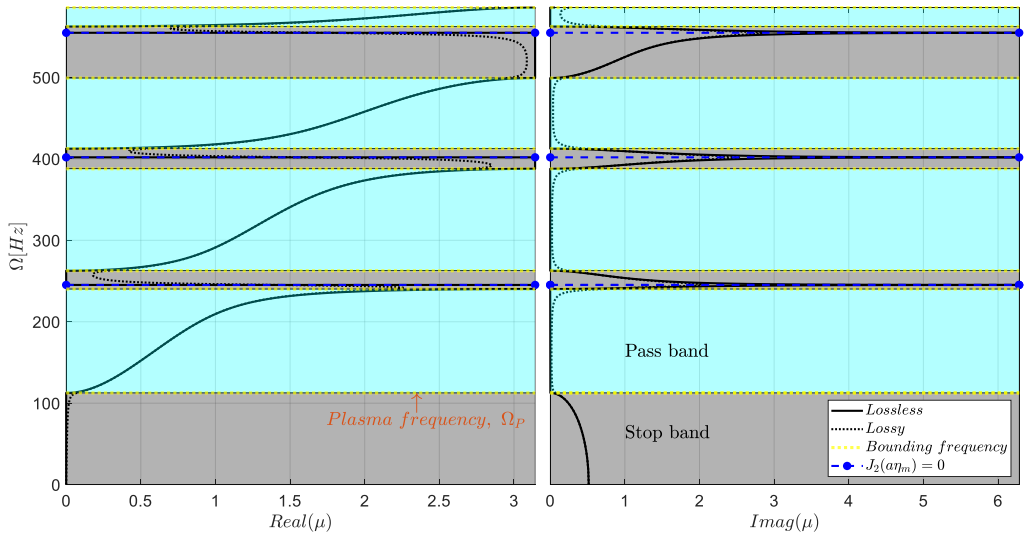

**Figure 2**: The four leading bandgaps of a typical vibroacoustic metamaterial with infinite number of monolayered unit-cells ($c_m = 15[m/s]$ and $\Delta_1 = 0.05[m]$): Real (left) and imaginary (right) parts of the propagation constant, $\mu$, where the bandgaps are marked by gray and the passbands by blue; the natural frequencies of the in-air membrane, satisfying $J_2(a\eta_m) = 0$, are shown by dashed horizontal lines with solid circles marking their edges.

A typical dispersion relation of the infinite vibroacoustic metamaterial is depicted in Fig. 2, where the four leading bandgaps are presented. Here the propagation constant, $\mu$, was computed using (26), the cut-on and cut-off frequencies (bounding frequencies) of the bands by (27), and the natural frequencies of the membrane interacting with the air in the cavity as roots of $J_2(a\eta_m) = 0$. By denoting the $n^{th}$ root of the second-order Bessel function of the first kind by $\gamma_2^{(n)}$, the $n^{th}$ natural frequency of the membrane, for the lossless and lossy cases, can be explicitly expressed as,

$$\Omega_{m,lossless}^{(n)} = \frac{c_m}{a}\gamma_2^{(n)}, \quad \Omega_{m,lossy}^{(n)} = \frac{i\lambda_m}{2\rho_m h_m} + \sqrt{\left(\frac{c_m}{a}\gamma_2^{(n)}\right)^2 - \left(\frac{\lambda_m}{2\rho_m h_m}\right)^2} \quad (31)$$

These relations provide countable infinities of natural frequencies for the in-air membrane.

Note that the fact that in the passbands, the propagation constant $\mu$ depends nonlinearly on frequency, indicates this is a highly dispersive vibroacoustic metamaterial. Moreover, while the standard Bragg-diffraction mechanism naturally generates bandgaps through spatial periodicity, these bandgaps predominantly occur at higher frequencies; as an indication, the frequency associated with first-order simple Bragg-diffraction is $\Omega_{Bragg}^{(1)} = \frac{1}{2\pi}\frac{\pi c_a}{\Delta_1} = 3,430[Hz]$. Consequently, the bandgaps observed in Fig. 2 cannot be attributed to the simple Bragg-diffraction mechanism alone. Indeed, the membrane inside each unit-cell acts as an internal local resonator which has the capability of storing the sound energy temporarily and suppressing sound transmission through the duct. Consequently, local resonance bandgaps emerge near each of the membrane's natural frequencies. This observation is consistent with the second, third, and fourth bandgaps illustrated in Fig. 2, with each bandgap corresponding to a distinct in-air membrane natural frequency.

Conventional discrete internal local resonators in phononic lattices, typically modeled as spring-mass subsystems, offer only a limited number of new bandgaps, equal to the number of local resonances. This constraint makes them impractical for applications requiring multi-frequency vibration absorption and limits their effectiveness in interacting with the high-frequency Bragg-diffraction bandgaps. However, the membrane local resonator supports an infinite number of local resonances and, consequently, bandgaps, thereby enabling broadband performance due to their interaction with Bragg bandgaps. In addition, unlike Bragg-diffraction bandgaps, these local resonance bandgaps are independent of the spatial periodic arrangement of the unit-cells and are solely dictated by membrane properties, as indicated in equation (31). As a result, local resonance sub-wavelength bandgaps are generated, overcoming limitations of phononic crystals reliant on Bragg scattering mechanisms, and offering *an additional way for manipulating (tailoring) acoustic waves at low-frequency ranges where Bragg scattering is typically ineffective*. Nevertheless, the local resonance bandgaps in acoustic metamaterials are typically rather narrow. However, in the vibroacoustic metamaterial considered these local resonance bandgaps become wider with increasing frequency (see Fig. 2). This phenomenon stems from the interplay between the local resonances and the Bragg-diffraction bandgaps. In fact, the higher the local resonance of the membrane, the more pronounced the interaction with Bragg-diffraction bandgaps becomes, resulting in wider bandgaps. Mathematically, the expressions for the Type II bandgaps in (29) show that the effect of cavity depth on the bandgap widths becomes significant at higher frequencies.

Besides the Bragg-diffraction and local resonance bandgap formation mechanisms, Fig. 2 reveals a distinct third mechanism for wave tailoring in the vibroacoustic phononic metamaterial. In particular, *an acoustic equivalence of plasma oscillation is observed in the* *first bandgap*, where the propagation of waves in the medium is completely prohibited in the frequency range $0 < \Omega < \Omega_P$. Here $\Omega_P$ is referred to as the *plasma frequency* and is computed by the first root of the Type II category of bandgaps in (29). This bandgap associated with the plasma-like phenomenon is referred to as *plasma bandgap*. This phenomenon is attributed to the negative effective density behavior exhibited by the membrane within the unit-cell, resulting in high reflectivity below its characteristic plasma frequency $\Omega_P$. Such plasma-like behavior was experimentally observed in [24] where a finite one-dimensional acoustic metamaterial with an array of elastic membranes was considered. In that study, the experimental findings were explained using an approximate theory based on deriving the effective density and bulk modulus of the considered metamaterial, whereas in the present study exact analytical expressions are provided.

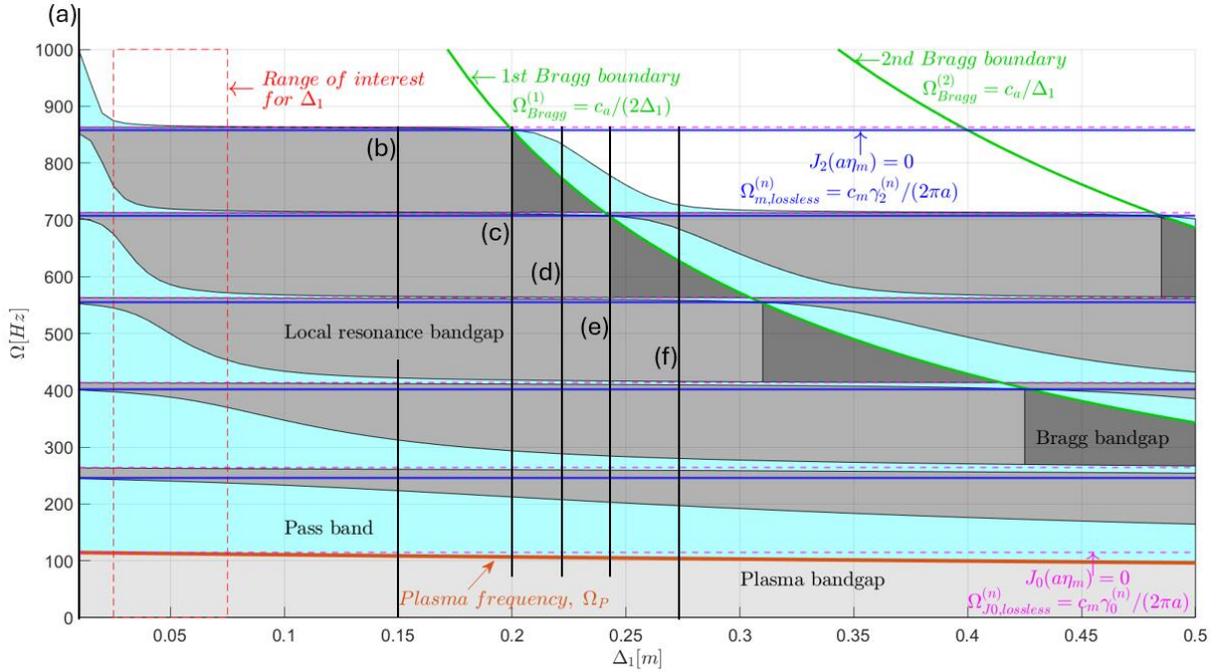

**Figure 3**: The six leading bandgaps of the vibroacoustic metamaterial with infinite number of monolayered unit-cells for $c_m = 15 [m/s]$ the varying cavity depth $\Delta_1$: Solid green curves represent the Bragg frequency boundaries, solid blue horizontal lines the lossless (undamped) in-air membrane natural frequencies, dashed magenta horizontal lines the in-vacuum membrane natural frequencies, and the orange curve the plasma frequency; the solid black vertical lines labeled (a) to (f) are selected cavity depths used to illustrate efficacy for wave tailoring of the metamaterial in Fig. 4.

Fig. 3 depicts the influence of cavity depth $\Delta_1$, i.e., unit-cell length, on the six leading bandgaps, where for clearer physical insight, the undamped membrane was considered, with fixed properties and leading five membrane natural frequencies, given by $\Omega_{m,lossless}^{(n)} = \frac{c_m}{2\pi a}\gamma_2^{(n)} [Hz]$ (represented by solid blue horizontal lines), where $J_2(\gamma_2^{(n)}) = 0$. Note that regardless of the cavity depth $\Delta_1$, local resonance bandgaps always occur in the vicinity of the membrane natural frequencies, $\Omega_{m,lossless}^{(n)}$, whereas first- and second-order simple Bragg-diffraction contours – derived from Type I bandgaps in (29), are depicted by the green curves.

The cut-off frequency $\Omega_P$ of the plasma bandgap shows weak dependence on the cavity depth. Moreover, applying a small-depth approximation based on Type II category in (29), the plasma frequency can be explicitly approximated as,

$$\Omega_P = \Omega_{Plasma}^{cut-off} = \frac{c_m}{2\pi a}\gamma_0^{(1)}[Hz] \approx 2.4048\frac{c_m}{2\pi a}[Hz], \quad \Delta_1 \ll 2c_a/\Omega \quad (32)$$

where $\gamma_0^{(1)}$ denotes the first root of $J_0\left(\gamma_0^{(1)}\right) = 0$. To the best of our knowledge, this is the first analytical approximation for the plasma frequency reported in the literature within practical unit-cell lengths, satisfying $\Delta_1 \ll 2c_a/\Omega$, for the monolayered vibroacoustic phononic metamaterial considered herein. Hence, the plasma frequency is found to be solely related to the membrane (internal resonator). Consequently, *within the monolayer vibroacoustic unit-cell, the membrane behaves as a mechanical high-pass filter, which permits the passage of waves with frequencies higher than the plasma cutoff frequency ($\Omega_P$) while attenuating waves with frequencies lower than $\Omega_P$*. Expanding the unit-cell length beyond the practical range causes a nonlinear decrease in the plasma frequency, as described in (29), exhibiting a softening effect, and the plasma frequency is computed as the first root of Type II bounding frequencies in (29).

Let $\Delta_{nl}$ denote a characteristic unit-cell length, where the $n^{th}$ local resonance frequency coincides with the $l^{th}$ Bragg frequency, defined as:

$$\Delta_{nl} = \frac{l\pi a c_a}{c_m \gamma_2^{(n)}} \quad (33)$$

For practical unit-cell lengths it holds that $\Delta_1 \ll 2c_a/\Omega$ and lengths slightly larger than the characteristic length $\Delta_{nl}$ (i.e., immediately following the interaction between the local resonance and Bragg bandgaps), the cut-off frequency of the $n^{th}$ local resonance bandgap remains unchanged with respect to the cavity depth $\Delta_1$, and is expressed as:

$$\Omega_{Local-res(n)}^{cut-off} = \frac{c_m}{2\pi a}\gamma_0^{(n+1)}[Hz], \quad (34)$$

where $\Delta_1 \ll 2c_a/\Omega$ or $0 < \Delta_1 - \Delta_{nm} \ll 2c_a/\Omega$.

At the same time, within these ranges of unit-cell lengths the cut-on frequency of the $n^{th}$ local resonance bandgap shows a pronounced dependence on the unit-cell length which is closely associated with the proximity to the Bragg-local resonance interaction point (i.e., the value of $\Delta_{nl}$). The smaller $\Delta_{nl}$ is, the sharper becomes the transition of the cut-on frequency of the $n^{th}$ local resonance bandgap, and, as it approaches the corresponding characteristic length, it asymptotically converges to the value:

$$\Omega_{Local-res(n)}^{cut-on} = \frac{c_m}{2\pi a}\gamma_0^{(n)}[Hz], \quad (35)$$

where $0 < \Delta_{nl} - \Delta_1 \ll 2c_a/\Omega$.

Let $W_{Local-res(n)}^{min}$ and $W_{Local-res(n)}^{max}$ denote the minimal and maximal widths of the $n^{th}$ local resonance bandgap, respectively. It is evident from Fig. 3 that the width of the $n^{th}$ local resonance bandgap evolves from $W_{Local-res(n)}^{min} \rightarrow W_{Local-res(n)}^{max}$ along the characteristic length $\Delta_{nl}$. These minimal and maximal widths are defined as:

$$W_{Local-res(n)}^{min} = \frac{c_m}{2\pi a}\left(\gamma_0^{(n+1)} - \gamma_2^{(n)}\right)[Hz]$$
$$W_{Local-res(n)}^{max} = \frac{c_m}{2\pi a}\left(\gamma_2^{(n)} - \gamma_0^{(n)}\right)[Hz] \quad (36)$$

The minimal width is obtained either near the zero length of the unit-cell or immediately after crossing the corresponding interaction Bragg-local resonance point. Conversely, the $n^{th}$ local resonance bandgap becomes broadest just before interacting with the Bragg mechanism. Therefore, *the broadening mechanism of the local resonance bandgap width becomes more efficient near the interaction point with Bragg scattering*. However, achieving proximity to Bragg scattering for low-frequency wave tailoring necessitates large and impractical lengths of unit-cells.

On the other hand, regardless of the membrane properties, the cut-off frequency of the $l^{th}$ Bragg bandgap is determined by the metamaterial periodicity, as follows:

$$\Omega_{Bragg(l)}^{cut-off} = \frac{lc_a}{2\Delta_1} [Hz] \tag{37}$$

Moreover, the cut-on frequency of the $l^{th}$ Bragg bandgap is independent of the unit-cell length and is determined by,

$$\Omega_{Bragg(l)}^{cut-on} = \frac{c_m}{2\pi a} \gamma_0^{(n)} [Hz] \tag{38}$$

where $\min_{n \in \mathbb{N}} \left\{ \frac{lc_a}{2\Delta_1} - \frac{c_m}{2\pi a} \gamma_0^{(n)} : \frac{lc_a}{2\Delta_1} > \frac{c_m}{2\pi a} \gamma_0^{(n)} \right\}$.

This Bragg cut-on frequency is achieved by expanding the Type II category equation (29) around the Bragg frequency. Combining (37) and (38) explicitly provides the width of the $l^{th}$ Bragg bandgap $W_{gap(l)}^{Bragg}$:

$$W_{gap(l)}^{Bragg} = \frac{lc_a}{2\Delta_1} - \frac{c_m}{2\pi a} \gamma_0^{(n)} [Hz] \tag{39}$$

where $\min_{n \in \mathbb{N}} \left\{ \frac{lc_a}{2\Delta_1} - \frac{c_m}{2\pi a} \gamma_0^{(n)} : \frac{lc_a}{2\Delta_1} > \frac{c_m}{2\pi a} \gamma_0^{(n)} \right\}$.

While the cut-off frequency of the Bragg bandgap depends solely on the spatial periodicity of the system, the cut-on frequency, conversely, relies solely on the local internal resonator (i.e., the membrane). As a result, the width of the Bragg bandgap is influenced by both the periodicity and the local resonator parameters. The precise relationship is described in (39). Therefore, in addition to adjusting the local resonance bandgaps through the introduction of the membrane, this also enables customization of the width of the Bragg bandgaps.

It is worth noting that the cut-on frequency of the $l^{th}$ Bragg bandgap becomes the cut-off frequency of the $n^{th}$ local resonance bandgap after crossing the corresponding characteristic length $\Delta_{nl}$, i.e., the interaction point between the $l^{th}$ Bragg bandgap and the $n^{th}$ local resonance bandgap. *A very narrow passband separates the $n^{th}$ local resonance bandgap and the $l^{th}$ Bragg bandgap before their interaction. However, after the interaction point, the separating passband broadens.* Consequently, the Bragg and local resonance bandgaps are always isolated by a passing band and cannot be hosted within a single bandgap. It follows that *the emergence of narrow and broad passing bands resulting from the interaction between local resonance and Bragg bandgaps facilitates the design and comprehension of acoustic wave band-pass filters for passive wave manipulation and tailoring.*

Focusing on a practical range of unit-cell lengths (satisfying $\Delta_1 \ll 2c_a/\Omega$, e.g., $2.5 \leq \Delta_1 \leq 7.5 [cm]$, as highlighted in the low-frequency dashed subregion in Fig. 3, it becomes evident that the local resonance and Bragg mechanisms for the formation of bandgaps remain isolated at low frequencies. This results in very limited controllability of passive wave tailoring and restricts the practical application of this type of vibroacoustic monolayered phononic metamaterials at low frequencies. Particularly, the first two pass bands are highly robust for changes in cavity depth within the practical length of the unit-cell.

To examine the wave attenuation performance within the bandgaps, Fig. 4 illustrates the imaginary part of the propagation constant, $Imag(\mu)$, for various unit-cell lengths ranging from $1[cm]$ to $27.5[cm]$. The aim here is to capture the attenuation characteristics of each bandgap mechanism and observe how the attenuation performance is influenced by the interaction between these mechanisms. This is achieved by changing the cavity depth and thus shifting the location of the Bragg bandgaps relative to the local resonance bandgaps. For small cavity depths, the local resonance bandgaps are isolated from the Bragg bandgaps (see Fig. 4a), and thus no interaction is observed. Consequently, these local resonance bandgaps are very narrow yet exhibit strong attenuation compared to the plasma bandgaps. As we increase the cavity length, according to (37) the Bragg bandgaps shift closer to the local resonance bandgaps (see Fig. 4b), leading to a notable change in the wave attenuation performance. Specifically, the local resonance bandgaps become very broad with asymmetric behavior, reaching maximum attenuation magnitude at the membrane natural frequencies. Furthermore, these local resonance bandgaps draw closer to each other as they approach the Bragg bandgaps but do not merge into a single bandgap; instead, they remain separated by narrow passbands, which become narrower as they approach the Bragg bandgap.

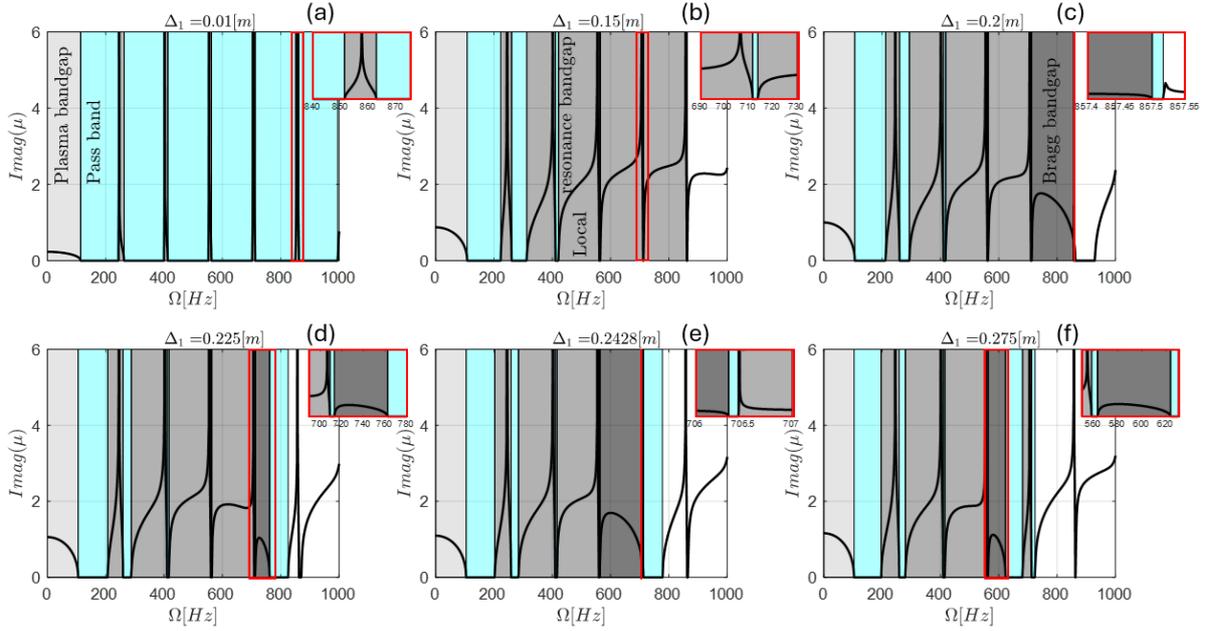

**Figure 4**: Capacity for passive wave attenuation as quantified by $Imag(\mu)$ for the six leading bandgaps at various cavity lengths, with (a) - (f) corresponding to the vertical lines in Fig. 3: The insets provide zoom-in views of the regions highlighted by the dashed boxes to highlight the existence of ultra-narrow passbands, with blue denoting passbands, light gray plasma bandgaps, medium gray local resonance bandgaps, and heavy gray Bragg bandgaps.

Fig. 4c illustrates the wave attenuation performance when the fifth natural frequency of the membrane (local resonance) coincides with the first Bragg cut-off frequency, corresponding to a cavity length equaling the characteristic length $\Delta_{51}$. In this scenario, the attenuation magnitude of the Bragg bandgap increases, and the widths of both the Bragg and nearby local resonance bandgaps reach their maximum values. Due to overlapping between the Bragg and local resonance bandgaps, ultra-narrow passbands are created to separate the bandgaps, thus preventing the formation of a single broad bandgap. It is noteworthy that the first passband remains largely unaffected as it is situated far from the interaction between the fifth local resonance and the first Bragg bandgap. As the unit-cell length increases further, the Bragg cut-on frequency remains unchanged, while its cut-off frequency decreases. This results in a narrowing of the Bragg bandgap width, accompanied by a decrease in its wave attenuation capacity compared to the overlapping with the local resonance scenario, as shown in Fig. 4d.

Such narrowing of the Bragg bandgap results in an expansion of the subsequent passband. As the cavity length is gradually increased from the characteristic length $\Delta_{51}$ to $\Delta_{41}$ (where the fourth local resonance bandgap overlaps with the first Bragg bandgap), the Bragg bandgap continues to shrink until reaching the point $\Delta_1 = \Delta_{41}$. At this point, the Bragg bandgap flips around its original cut-on frequency, which then becomes the new cut-off frequency, while the new cut-on frequency shifts to a much lower frequency. This creates a new broad Bragg bandgap characterized by high attenuation, as shown in Fig. 4e. As the cavity length is increased further within the range $\Delta_{41} < \Delta_1 < \Delta_{31}$, the attenuation capacity of the Bragg bandgap decreases, and its cut-on frequency remains unaffected while its cut-off frequency gradually decreases, leading to a narrowing of its width (see Fig. 4f).

In all scenarios illustrated in Figure 4, there is no merging of the local resonance and Bragg bandgaps into a single bandgap. Instead, the bandgaps remain distinct with narrow passbands separating them. These narrow passbands are bounded by a local resonance bandgap and a Bragg bandgap, with significant attenuation performance obtained around the edge of the narrow passband due to the local resonance of the membrane. Further characteristics of these narrow passbands will be demonstrated in the next section for the finite periodic lattice. Lastly, we emphasize that, even with the use of relatively large unit-cell lengths, such as $\Delta_1 = 27.5 [cm]$, the ability of the vibroacoustic monolayered mrtamaterial to control and engineer the first band is rather limited.

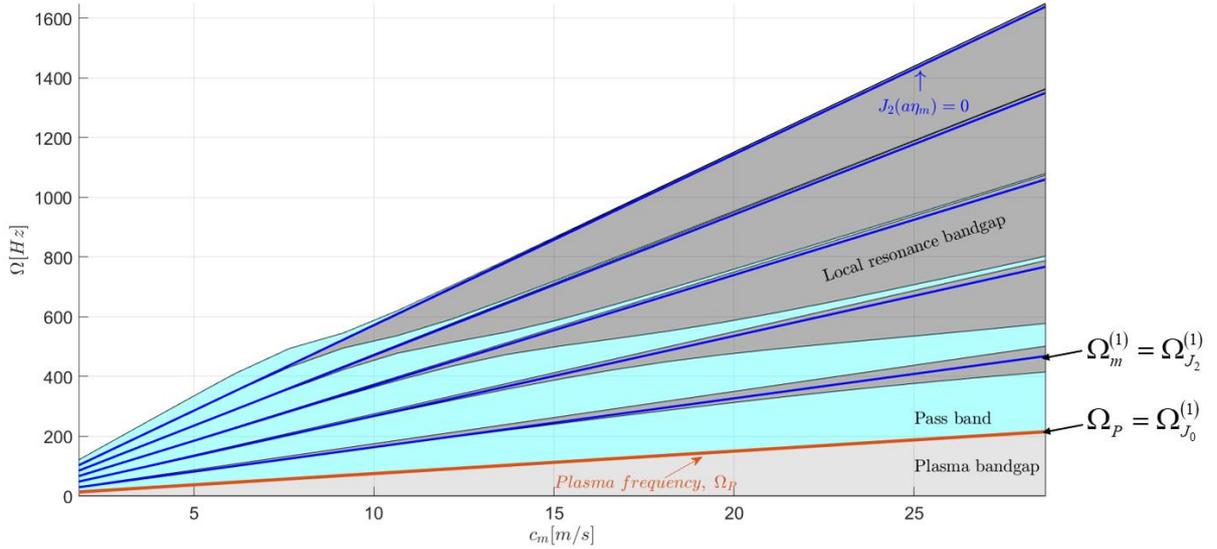

**Figure 5**: Evolution of the six leading bandgaps in the infinite monolayered vibroacoustic metamaterial with $\Delta_1 = 0.05 [m]$ as the speed of elastic wave propagation in the membrane ($c_m$) varies; blue denotes passbands, light gray the plasma bandgap, and medium gray local resonance bandgaps.

Considering a unit-cell with cavity length $\Delta_1 = 5 [cm]$, the role of the elastic wave speed in the membrane ($c_m$) on the topology of the six leading bandgaps is illustrated in Fig. 5. The plasma frequency is accurately described by the $1^{st}$ root of the zero-order Bessel function of the first kind, $J_0$, as shown in (32). Additionally, the cut-off frequency of the $n^{th}$ local resonance bandgap is related to the $(n+1)^{th}$ root of $J_0$ as described in (34). Since the standard first Bragg bandgap is located at the relatively frequency $\Omega_{Bragg}^{(1)} = \frac{1}{2\pi}\frac{\pi c_a}{\Delta_1} = 3,430 [Hz]$, only high-frequency bands are affected due to their proximity to the first Bragg bandgap. As the wave speed $c_m$ increases, the band structure simply exhibits a linear stretching of the cut-off frequencies of the bandgaps, while the cut-on frequencies behave in a nonlinear manner. Affecting though the first (lowest frequency) passband though is not possible for the monolayered vibroacoustic metamaterial considered.

## 3. The finite periodic lattice with N unit-cells

### 3.1 The global transfer matrix

As all metamaterials are of finite spatial extent, it is necessary to consider the finite phononic vibroacoustic metamaterial composed of only $N$ monolayered unit-cells, with schematic side-view in Fig. 6. The acoustic state vector of the lattice at the left and right boundaries are given by $\left[\frac{p_{L_1}}{\rho_a c_a} \; v_{L_1}\right]^T$ and $\left[\frac{p_{R_N}}{\rho_a c_a} \; v_{R_N}\right]^T$, respectively. Utilizing the local transfer matrix $\boldsymbol{T}$ representation (16), the *global transfer matrix* connecting the left and right state vectors is:

$$\begin{bmatrix} -\dfrac{p_{R_N}}{\rho_a c_a} \\ v_{R_N} \end{bmatrix} = \boldsymbol{T}^N \begin{bmatrix} \dfrac{p_{L_1}}{\rho_a c_a} \\ v_{L_1} \end{bmatrix} \tag{40}$$

Since the local transfer matrix, $\boldsymbol{T}$, is a $(2 \times 2)$ matrix, using the Cayley–Hamilton theorem, one can express its $N^{th}$ power, i.e., the global transfer matrix, in the form:

$$\boldsymbol{T}^N = \alpha \boldsymbol{I} + \beta \boldsymbol{T} \tag{41}$$

where $\boldsymbol{I}$ is the $(2 \times 2)$ identity matrix, and $\alpha$ and $\beta$ are real scalars. According to the Cayley–Hamilton theorem, the eigenvalues of the unimodular matrix $\boldsymbol{T}$, given by $\Lambda_{1,2} = e^{\pm i\mu}$ must also satisfy equation (41). Therefore, one obtains the following two relations:

$$\begin{aligned} e^{iN\mu} &= \alpha + \beta e^{i\mu} \\ e^{-iN\mu} &= \alpha + \beta e^{-i\mu} \end{aligned} \tag{42}$$

Solving for the unknowns $\alpha$ and $\beta$ and substituting into (41), the global transfer matrix $\boldsymbol{T}^N$ can be explicitly expressed as,

$$\boldsymbol{T}^N = \begin{bmatrix} \dfrac{T_{11}\sin(N\mu) - \sin((N-1)\mu)}{\sin(\mu)} & \dfrac{T_{12}\sin(N\mu)}{\sin(\mu)} \\ \dfrac{T_{21}\sin(N\mu)}{\sin(\mu)} & \dfrac{T_{22}\sin(N\mu) - \sin((N-1)\mu)}{\sin(\mu)} \end{bmatrix} \tag{43}$$

where $T_{11}, T_{12}, T_{21}$ and $T_{22}$ are the components of the local transfer matrix $\boldsymbol{T}$ as defined in (16), and $\mu$ is the propagation constant given in (26). Note that up to this point we did not account for the specific boundary conditions of the system at the left and right boundaries. These will be imposed next to compute the $N$ vibroacoustic modes.

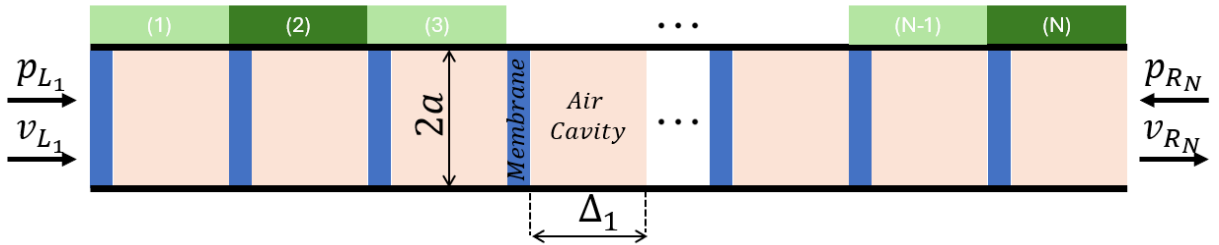

**Figure 6**: Schematic representation of a finite phononic vibroacoustic metamaterial composed of $N$ monolayered unit-cells and arbitrary boundary conditions.

## 3.2. The vibroacoustic modes

Applying different types of boundary conditions in terms of acoustic pressure and velocity, to the left and right boundaries of the finite vibroacoustic metamaterial yields characteristic equations for the corresponding natural frequencies. Without loss of generality, the case of free-free edges, i.e., $p_{L_1} = p_{R_N} = 0$, will be considered, which, when introduced into the global transfer matrix (40) and taking into account (43), yields the following characteristic equations:

$$T_{12}(\Omega) = 0 \text{ or } \sin(N\mu(\Omega)) = 0 \qquad (44)$$

These expressions are implicit functions of the frequency $\Omega$, and need to be solved for the natural frequencies of a finite lattice with free edges on both sides. The first expression, $T_{12}(\Omega) = 0$, does not depend on the number of unit-cells, $N$, and represents the natural frequencies of a single unit-cell with both edges free; the corresponding resonances occur either within the bandgaps or at the boundaries of passbands. The second expression, however, $\sin(N\mu(\Omega)) = 0$, provides exactly $(N-1)$ natural frequencies located within each passband of the finite phononic metamaterial.

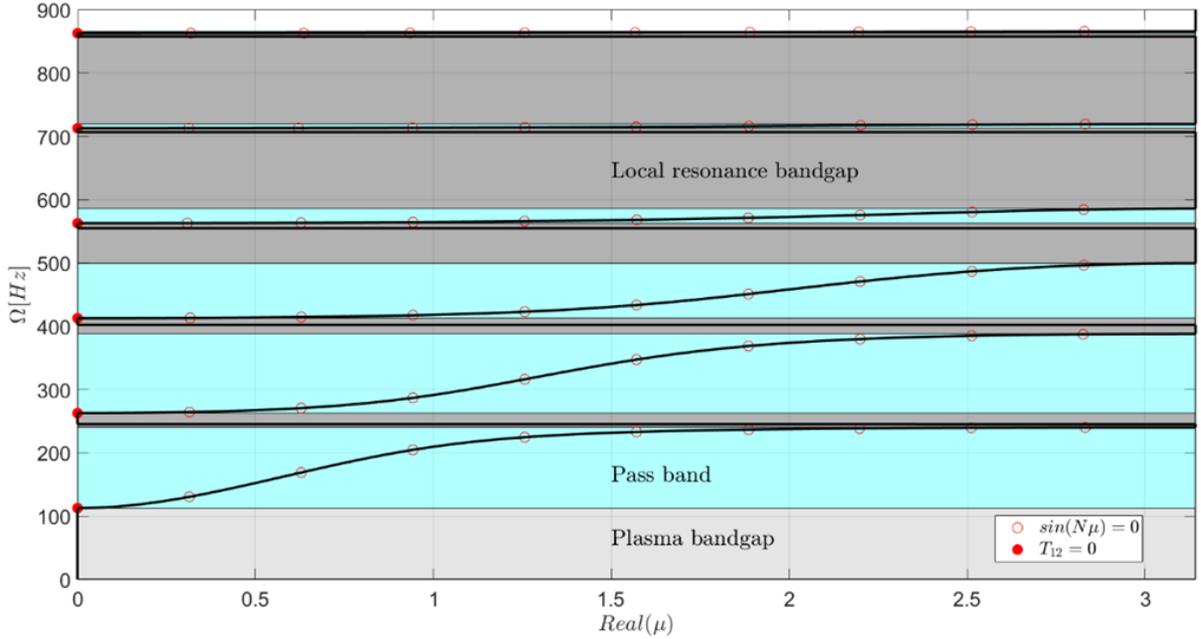

**Figure 7**: The projection of the natural frequencies (red circles) for a free–free monolayered vibroacoustic phononic metamaterial with $N = 10$ unit-cells on the dispersion branches of the passbands of the corresponding infinite system for $c_m = 15[m/s]$ and $\Delta_1 = 0.05[m]$.

The plot in Fig. 7 illustrates the natural frequencies (marked by red circles) of the free-free *finite* system composed of $N = 10$ unit-cells, projected onto the dispersion branches of the six leading passbands of the corresponding *infinite* system. The red filled circles represent the roots of $T_{12}(\Omega) = 0$, which remain consistent regardless of the number of unit-cells, being positioned at the bounding frequencies of the passbands. The red unfilled circles denote the $(N-1)$ roots of $\sin(N\mu(\Omega)) = 0$, that are equally distributed within each passband, dividing each dispersion branch inside each passband into $N$ partitions. It is seen that the narrow higher frequency passbands become densely spaced, creating a transparency window within the typically impermissible band. This transparency window is induced by the interplay of local resonances and Bragg bandgaps, and it appears to be an acoustic analogue of the well-known electromagnetically induced transparency caused by quantum interference [42].

An alternative physical interpretation of the natural frequencies can be ascribed to the scattering coefficients. Fig. 8 shows the transmission, reflection, and absorption coefficients of the finite phononic metamaterial composed of $N = 10$ unit-cells. These scattering coefficients of the finite lattice were obtained by introducing the global transfer matrix components (43) into the scattering matrix (19), i.e., by replacing the local transfer matrix components by the global transfer matrix components. The natural frequencies are highlighted by dashed-black columns in Fig. 8.

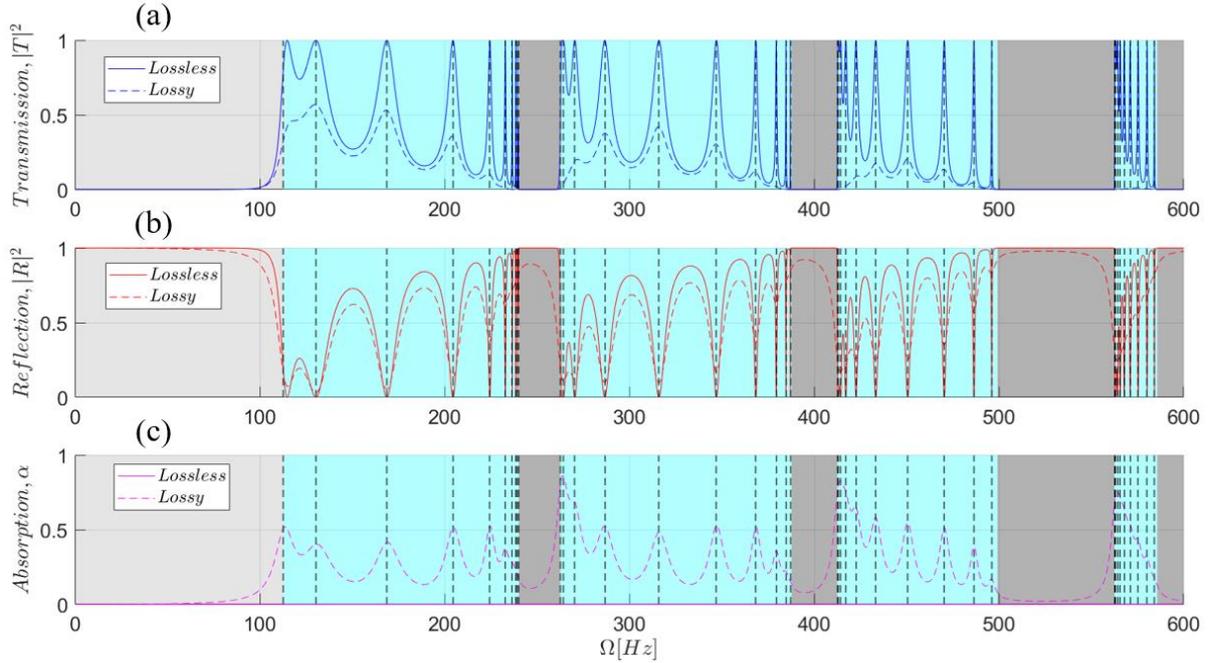

**Figure 8**: Scattering coefficients of the finite vibroacoustic metamaterial composed of $N = 10$ unit-cells: Transmission (a), reflection (b), and absorption (c), for $c_m = 15[m/s]$ and $\Delta_1 = 0.05[m]$.

Even though the finite system was constructed from a relatively small number of unit-cells, namely $N = 10$, the appearance of completely reflective (characterized with unit reflection and zero transmission) frequency bands is evident in the results of Fig. 8. These bands correspond to the bandgaps of the system with infinite number of unit-cells. In terms of scattering coefficients, these "forbidden bands" are separated by frequency bands where the transmission coefficient reaches peaks of unity, indicating resonant transmission, that is, perfect transmission where the undamped finite lattice becomes completely transparent. Each pass band contains $N$ peaks of unit transmission occurring precisely at the $N$ discrete natural frequencies. Therefore, only acoustic waves corresponding to the collective modes of the undamped finite system can freely propagate through the lattice. However, the presence of losses in the lattice reduces transmission and causes absorption peaks at the same natural frequencies of the finite system, effectively eliminating these collective modes.

The presence of non-perfect transmission within the passbands can be attributed to the boundary effects of the finite system, which generate reflected waves interacting with the incident ones. By combining the left and right incident scattering states and taking the limit as $N \to \infty$, *Bloch states* can be generated. These states propagate through the infinite periodic lattice without experiencing any reflections at the interfaces between adjacent vibroacoustic unit-cells. Moreover, within the narrower passbands depicted in Fig. 8, it is observed that the resonant transmissions are closely spaced. In contrast to the broader passbands, the transmission coefficient exhibits minimal reduction between the resonant transmissions, resulting in the passband appearing *nearly transparent*. Such acoustical transparency is caused by the interaction between the local resonance and Bragg bandgaps as described in [42].

## 4. Concluding remarks

In this work we studied a vibroacoustic phononic metamaterial consisting of repetitive monolayered unit-cells, that is, composed of a single elastic membrane interacting with an adjacent air cavity. The resulting fluid-structure interaction generates interesting vibroacoustic phenomena, with the membrane acting as an internal resonator within the unit-cell and enabling passive wave tailoring. For the infinite unbounded system (i.e., the one with an infinite number of unit-cells), the Bloch-Floquet theorem enables the study of wave propagation and attenuation in the system based on eigenvalue analysis of a single unit-cell. In that regard, our focus was on the low-frequency range, $0 < \Omega < \frac{3.8317}{2\pi a} c_a = 4183\ [Hz]$ (where $a = 0.05\ [m]$, $c_a = 343\ [m/s]$). By assuming axisymmetric longitudinal modes for the pressure inside the air cavity and axisymmetric transverse elastic modes for the membrane, we obtained relatively simple, exact closed form solutions for the vibroacoustic problem, eliminating the need of infinite series as is the current state-of-the-art. Indeed, unlike traditional analytical methods, which typically represent the axisymmetric transverse deflection of the membrane using an infinite series of eigenfunctions (normal modes), we derived a single-term exact solution for the forced-damped membrane that fully incorporates the effects of the air cavity.

Using these explicit solutions, we constructed a unimodular $(2 \times 2)$ frequency-dependent local transfer matrix, which, in turn, enabled the analytical construction of the corresponding local scattering matrix which fully characterizes the transmitted and reflected waves in the infinite metamaterial system, and also accounts for absorption when losses (damping effects) are considered. Thereafter, the transcendental characteristic equation for determining the bounding frequencies of the passbands of the metamaterial provided a comprehensive understanding of the mechanisms governing the vibroacoustic interactions within the unit-cells which generated either bandgaps for negative interference between reflected and transmitted waves at unit-cell interfaces, or passbands for constructive interference (and appropriate phase differences) between these waves.

Hence, three distinct mechanisms for bandgap formation were identified in the vibroacoutic metamaterial of infinite extent, namely, Bragg, local resonance, and plasma bandgaps. These bandgaps define frequency ranges where sound propagation within the periodic lattice is prohibited. More interestingly, the interactions between these bandgap mechanisms and their influence on each other was shown to provide passive tunability in the vibroacoustics, e.g., by varying the Bragg bandgap placement. This is achieved by varying the cavity depth (i.e., the spatial periodicity of the metamaterial) while keeping the membrane properties fixed and thus shifting the position of the Bragg bandgaps relative to the local resonance bandgaps. In addition, we reported a simple and accurate analytical approximation for the plasma frequency within practical unit-cell lengths ($\Delta_1 \leq 0.1\ [m]$), showing its exclusive dependence on the membrane properties. Minimal and maximal widths of the local resonance bandgaps, as well as limiting cases for the corresponding cut-on and cut-off frequencies were also derived. The results showed that the broadening mechanism of the local resonance bandgaps becomes more efficient near their interaction points with Bragg scattering bandgaps. The Bragg cut-off frequencies are solely dependent on the spatial periodicity of the metamaterial, whereas the Bragg cut-on frequency is determined only by the properties of the membrane (the internal resonator). Therefore, the membrane not only adjusts local resonance bandgaps but also allows for tailored customization of Bragg bandgap widths. Achieving proximity to Bragg scattering for low-frequency wave tailoring necessitates large and impractical lengths of unit-cells for the monolayered metamaterial system considered. We investigated the wave attenuation performance within the bandgaps for different cavity depths, encompassing scenarios where the local resonance and Bragg bandgaps are isolated, closely situated, overlapping, and post-interaction. In all these scenarios, there was no merging of the local resonance and Bragg bandgaps into a single bandgap. Instead, the bandgaps remained distinct with narrow passing bands separating them.

Lastly, for the finite metamaterial system with $N$ unit-cells, the $(2 \times 2)$ global transfer matrix was explicitly constructed. This approach facilitated deriving straightforward expressions for the natural frequencies under the assumption of both boundaries of the finite system being free, but similar analytical expressions hold for other simple types of boundary conditions (e.g., clamped). The relation between the natural frequencies of the system, the band structure, and the resonant transmission points was established. Within the narrow passbands resulting from the interaction between the local resonance and Bragg bandgaps, resonant transmissions were observed to be closely spaced, creating transparency windows. This vibroacoustic phenomenon bears resemblance to electromagnetically induced transparency caused by quantum interference.

The results uncover the presence of a plasma bandgap with a cut-on frequency of zero frequency. Consequently, the monolayered vibroacoustic metamaterial has demonstrated effectiveness in suppressing longitudinal vibrations within an exceptionally low frequency range up to the plasma frequency. This efficacy is intricately connected to the properties of the membrane within each unit-cell. However, we have also identified significant limitations in manipulating (tailoring) sound waves at low frequencies beyond the plasma frequency for unit-cell lengths satisfying $\Delta_1 \ll 2c_a/\Omega$ for the monolayered phononic system considered herein. This constrains the practical application of this hononic metamaterial at very low-frequency scenarios above the plasma frequency. Notably, the first two passbands were shown to be highly robust against variations in cavity length within the practical characteristic length of the unit-cell. However, the presented theoretical framework paves the way to engineer different types of vibroacoustic unit-cells to passively tailor and control the attenuation of low-frequency sound waves, e.g., using multiple differing membranes within the unit-cell; this is the topic of current research by the authors. Moreover, our findings are not limited to sound propagation, but apply for wave propagation in more general classes of elastic metamaterial systems where structural sound is of interest.

**Acknowledgements**

The authors are grateful to the Israeli Council of Higher Education (CHE-VATAT), and the Department of Mechanical Science and Engineering (MechSE) at the University of Illinois Urbana-Champaign for financial support.

**Data Availability Statement**

This manuscript has no associated data.

**Declarations Conflict of interest**

The authors declare that they have no conflict of interest.